# High-temperature interface superconductivity between metallic and insulating cuprates

A. Gozar[1], G. Logvenov[1], L. Fitting Kourkoutis[2], A. T. Bollinger[1], L. A. Giannuzzi[3], D. A. Muller[2], I. Bozovic[1]

[1]Brookhaven National Laboratory, Upton, New York 11973-5000, USA

[2]School of Applied and Engineering Physics, Cornell University, Ithaca, NY 14853, USA

[3]FEI Company, Hillsboro, OR 97124, USA

**High-temperature superconductivity confined to nanometer-size interfaces has been a long standing goal because of potential applications[1,2] and the opportunity to study quantum phenomena in reduced dimensions[3,4]. However, this is a challenging target: in conventional metals the high electron density restricts interface effects such as carrier depletion/accumulation to a region much narrower than the coherence length, the scale necessary for superconductivity to occur. In contrast, in copper oxides the carrier density is low while the critical temperature ($T_c$) is high and the coherence length very short; so, this provides a breakthrough opportunity - but at a price: the interface must be atomically perfect. Here we report on superconductivity in bilayers consisting of an insulator ($La_2CuO_4$) and a metal ($La_{1.55}Sr_{0.45}CuO_4$), neither of which is superconducting in isolation. However, in bilayers $T_c$ is either ~15 K or ~30 K, depending on the layering sequence. This highly robust phenomenon is confined within 2-3 nm from the interface. If such a bilayer is exposed to ozone, $T_c$ exceeds 50 K and this enhanced superconductivity is also shown to originate from the interface layer about 1-2 unit cell thick. Enhancement of $T_c$ in bilayer systems was observed previously[5] but the essential role of the interface was not recognized at the time. Our results demonstrate that engineering artificial heterostructures provides a novel, unconventional way to fabricate stable, quasi

**two-dimensional high $T_c$ phases and to significantly enhance superconducting properties in known or new superconductors.**

Typical approaches for the realization of quasi two-dimensional superconducting sheets rely on fabrication of an 'ultrathin' layer of a known superconductor[1,2]. Another route is to use hetero-interfaces. Superconductivity in the 0.2-6 K range was reported at the interface between two oxide insulators[6] or in superlattices where one[7] or both[8] components are semiconductors. The $La_{2-x}Sr_xCuO_4$ (LSCO) family is particularly attractive because these materials are solid solutions that can be doped over a broad range[9].

In our experiment, we have synthesized a large number (over 200) of single-phase, bilayer, and trilayer films with insulating (*I*), metallic (*M*) and superconducting (*S*) blocks in all combinations and of varying layer thickness (for the notation see the caption to Fig. 1). The films were grown in a unique atomic-layer-by-layer molecular beam epitaxy (ALL-MBE) system[10] that incorporates *in situ* state-of-the-art surface science tools such as time-of-flight ion scattering and recoil spectroscopy (TOF-ISARS) and reflection high-energy electron diffraction (RHEED). It enables synthesis of atomically smooth films as well as multilayers with perfect interfaces[5,11,12,13]. Typical surface roughness determined from atomic force microscopy (AFM) data is 0.2 – 0.5 nm, less than one unit cell (UC) which in LSCO is 1.3 nm. ALL-MBE provides for digital control of layer thickness, which we measure by counting the number of UCs. Maintaining atomic scale smoothness and digital layer-by-layer growth are both crucial for the results we discuss in the following.

The interface between the metallic and insulating materials is superconducting with high $T_c$ (see Fig. 1) and the deposition sequence matters. *M-S* bilayers show the highest critical temperature, $T_c \approx 50$ K. In contrast, in single-phase LSCO films which we have grown under the same conditions, the highest $T_c$ is about 40 K, similar to what is seen in bulk single crystals (ref. 9 and Supplementary Fig. 1). Hence, in *M-S* bilayers we see a large (up to 25%) relative $T_c$ enhancement. $T_c$'s around 50 K were observed previously in some LSCO films[14,15] and LSCO-LCO bilayers[5] but the locus of the highest $T_c$ has not been investigated. Below we show that in our *M-I* films enhanced superconductivity originates from and is restricted to a 1-2 UC thick interfacial layer. In retrospect, one would suppose that at least the bilayer result[5] was also an interface effect, a proposition that we confirmed, as discussed below.

To directly determine the length scale associated with interface superconductivity we synthesized a series of *M-I* and *I-M* structures with thick bottom layers ($\geq$ 30 UC) while the thickness of the top layer was increased digitally, one-half UC at a time (Fig. 2). The transport data show that the plateau values for superconductivity are reached after the thickness of the top layer is $\geq$ 2UCs, a value which sets the length scale for this interface phenomenon.

The $T_c$ enhancement in *M-S* bilayers triggers the intriguing question whether this enhancement is an interface phenomenon as suggested by several preliminary observations, see Supplementary Information. That this is the case is confirmed by the data on critical current density ($j_c$) determined from two-coil mutual inductance measurements[16-18] (see Fig. 3). The results indicate that the $T_c \approx$ 50 K in *M-S* structures is in fact confined to a very thin (1-2 UC thick) layer near the interface. The observed linear temperature dependence of $j_c$ in *S* films is expected theoretically in cuprates for the intrinsic critical current due to vortex-antivortex pair breaking or depinning in homogeneous samples[19], and it is observed experimentally in high-quality high-temperature superconductors (HTS) films and bulk single crystals[20]. In contrast, in *M-S* samples one can see a clear break near 40 K which separates two approximately linear regions with very different slopes.

This is what one expects from two superconducting sheets with different thickness and critical temperature, say $d_1$, $T_{c1}$ and $d_2$, $T_{c2}$, respectively. The breakdown into two such components (the dashed lines in Fig. 3) provides $T_{c1} \sim 40$ K and $T_{c2} \sim 50$ K. The low-temperature extrapolation of the critical current gives $d_1/d_2 \sim 20$. Since the total number of layers deposited was $d_1 + d_2 = 20$ UC, one obtains $d_2 \sim 1$ UC. This length scale is quantitatively consistent with the independent measurements of resistivity in *M-S* bilayers as a function of top layer thicknesses (see Fig. 2c). We performed similar mutual inductance measurements on the exact same bilayer sample (not deteriorated after seven years) studied in ref. 5 in which the bottom layer was optimally doped LSCO and the results were quite similar to the *M-S* case; this demonstrates that the previously reported $T_c$ enhancement was also an interface effect.

The issue of interface structure and possible impact of cation interdiffusion is discussed in Fig. 4. The microstructure of an *M-I* bilayer and its interfaces was analyzed using electron energy loss spectroscopy (EELS) in a scanning transmission electron microscope (STEM). An upper limit on the amount of chemical interdiffusion at the interfaces is obtained by recording the Lanthanum-$M_{4,5}$ EELS edges. The rms interface roughness, as determined by fitting error functions to the La profile, is $\sigma = 0.8 \pm 0.4$ nm at the substrate-*M* interface and $\sigma = 1.2 \pm 0.4$ nm (~ 1 UC) at the *M-I* interface, which sets an upper limit to any cation intermixing, see also Supplementary Fig. 8.

As an independent test of chemical variations across the interfaces, the changes in the Oxygen-K fine structure were analyzed using a principle-components analysis. The fraction of the component corresponding to the *M* layer is shown in Fig. 4(d), which again indicates an interface roughness less than 1 UC. Either interface was fully described by two components, leaving no significant residual after the fit suggesting that there is no substantial third, interfacial layer, at least on the scale of the interface roughness. Results obtained by several other surface sensitive probes like RHEED and TOF-ISARS as well as transport on *I-M-I* hetero-structures (see Supplementary Figs. 2, 3 and 4) support and are consistent with the chemically abrupt interfaces inferred from the STEM data. The experiments set an upper limit on possible cation interdiffusion to less than 1 UC and make the cation mixing scenario hard to reconcile quantitatively and qualitatively with our observations.

Other possible causes for the interface HTS are electronic reconstruction or oxygen non-stoichiometry. Experimental data show that charge depletion or accumulation is substantial across *M-I* and *I-M* interfaces[23] while such charge transfer is negligible when *M* is replaced by optimally doped LSCO, ref. 15. These findings are consistent with the doping dependence of the chemical potential in LSCO inferred from X-ray photoemission data[24]. Oxygen vacancies and interstitials are nevertheless additional factors that should be considered: they may account for the asymmetry between *M-I* and *I-M* structures and are essential for increased $T_c$ and stability of superconductivity in *M-S* bilayers, see section E of the Supplementary Information.

A remaining puzzle is the mechanism of relative $T_c$ enhancement in *M-S* bilayers. It is conceivable that structural aspects, such as disorder, play a crucial role. We may have realized the doping without disorder scenario[25] by introducing carriers via charge transfer and by (ordered) interstitial oxygen pinned near the interface. Another possibility is that the "intrinsic" $T_c$ in LSCO would be even higher were it not for some competing instability and that this other order parameter is suppressed in bilayers via the long-range strain and/or electrostatic effects. Finally, an interesting possibility is that pairing and/or coherence of electrons in one layer is enabled or enhanced by interactions originating in the neighbouring layer[26,27]. Deciphering this problem may open the path to even larger $T_c$ enhancement.

**Figure 1 The dependence of resistance on temperature for single-phase and bilayer films**. Notation used in text and figures: $I$ = $La_2CuO_4$, vacuum-annealed and insulating; $S$ = $La_2CuO_{4+\delta}$, oxygen-doped by annealing in ozone and superconducting; $M$ = $La_{1.55}Sr_{0.45}CuO_4$, overdoped and metallic but not superconducting. For bilayers, the first letter always denotes the layer next to the $LaSrAlO_4$ substrate. Panels (a) and (b): $R(T)$ for single-phase layers of $I$ (note the log scale) and $M$, respectively. Panel (c): R($T$) normalized to $T$ = 200 K for various bilayers. The typical values for the superconducting critical temperature ($T_c$) at the mid-point of the resistive transitions are: $T_c \approx$ 15 K in *I-M* and $T_c \approx$ 30 K in *M-I* structures. In *M-S* bilayers (four samples shown) $T_c \approx$ 50 K. In a few hundred single-phase films (doped by either oxygen or Sr) grown under the same conditions, $T_c$ never exceeded 40 K, the value marked by the arrow,

see Supplementary Fig. 1. The interface superconductivity is reproducible and stable in air on the scale of months in contrast to single-phase *S* films.

**Figure 2 The dependence on the layer thickness.** (a) Normalized resistance for several *I-M* bilayers where the thickness of the bottom *I* layer is fixed at 40 unit cells (UC), i.e., 52 nm, while the thickness of the *M* layer is varied as indicated. For a 0.5 UC thick *M* layer the sample is insulating while the 1.5 UC structure shows a metallic to insulating crossover near *T* = 75 K. Further increase of the thickness raises $T_c$ to a 15 K plateau. (b) The same for *M-I* bilayers with a 40 UC thick bottom *M* layer. Traces of superconductivity can be observed even when the bottom *M* layer is covered by only 0.5 UC (0.66 nm) thick *I* layer. When 1 UC of *I* covers the surface, the resistive transition is complete and $T_c$ > 10 K. On its own, this is a signature of virtually atomically perfect surfaces given that the resistance measurements were taken with the voltage probes 3 mm apart. (c) $T_c$ (defined as the midpoint of the resistive transition) as a function of the top layer thickness in *M-I*, *I-M* and *M-S* bilayers. The latter are structures obtained by annealing *M-I* bilayers in ozone atmosphere, the procedure that turns *I* films into *S* while having essentially no effect on *M*. The dashed lines are guides for the eye.

**Figure 3 Non-linear screening effects in a single-phase (*S*) film and a *M-S* bilayer.** (a) The dependence of the pick-up voltage on the current in the drive coil at several temperatures. At each temperature, a 'critical' value of the current in the drive coil, $I_{dc}$, corresponds to the onset of dissipation in the film and can be defined as the crossover point between a linear (*n* = 1) and a higher-power law ($n \approx 3$ at temperatures below 40 K) behaviour. In both samples the *S* layer is 20 UC thick. (b) The temperature dependence of $I_{dc}$ for an *S* film (filled diamonds) and an *M-S* bilayer (empty squares). The right scale shows the calculated peak value of the induced screening current density in superconducting films, see also Supplementary Fig. 5. Arrows denote the critical temperatures, $T_c$ = 33.2 K and $T_c$ = 51.6 K for the *S* and *M-S* samples respectively. The bilayer data can be well decomposed into two approximately linear contributions (dashed lines), corresponding to bulk and interface parts with $T_c \approx$ 40 K and $T_c \approx$ 50 K as shown in the lower left drawing. The inset shows the same data in reduced temperature units $T/T_c$. The magnitude of the estimated low-temperature critical current of the thin layer is in agreement with the value estimated from mutual inductance and transport measurements in *M-I* bilayers in which the HTS ($T_c$ = 30 K) sheet has a similar thickness.

**Figure 4 Scanning transmission electron microscopy and electron energy loss spectroscopy analysis of an *M-I* bilayer.** (a) Annular dark field image of the structure. A magnified image of the *M-I* interface, marked by arrows, is shown in the inset. (b) O-K EELS of the three oxides in the structure showing clear changes in the fine structure of the O-K edge. For LSCO an O-K edge pre-peak (circled) evolves for $x$ > 0 and scales with the doping level[21,22]. (c) The integrated La intensity across the bilayer. As expected, the La profile shows an increase in the La concentration from the substrate to the *M* layer and again from the *M* to the *I* layer. (d) Results of a principle-components analysis of the two interfaces. Here, the fraction of one of two components, corresponding to the O-K edge in $La_{1.55}Sr_{0.45}CuO_4$, is shown.

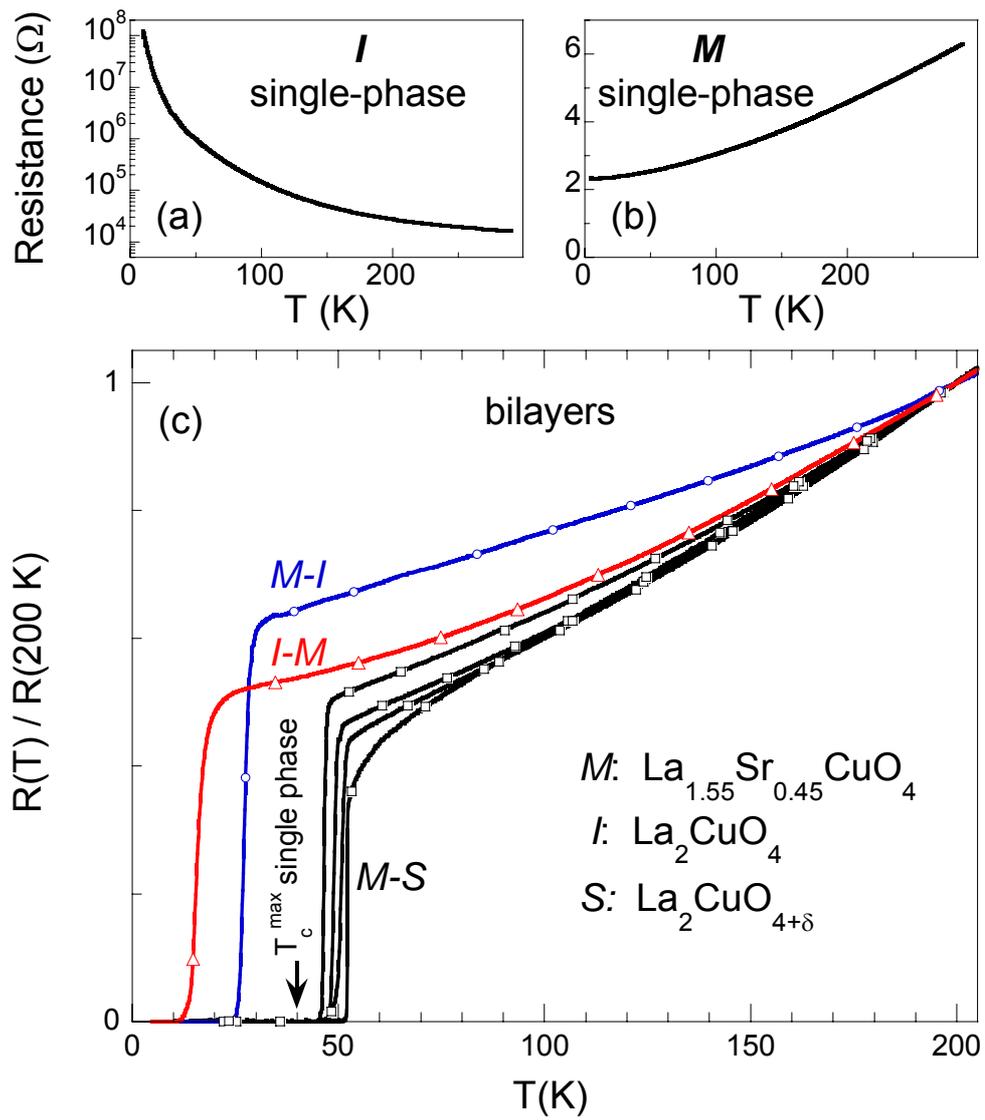

**Figure 1**

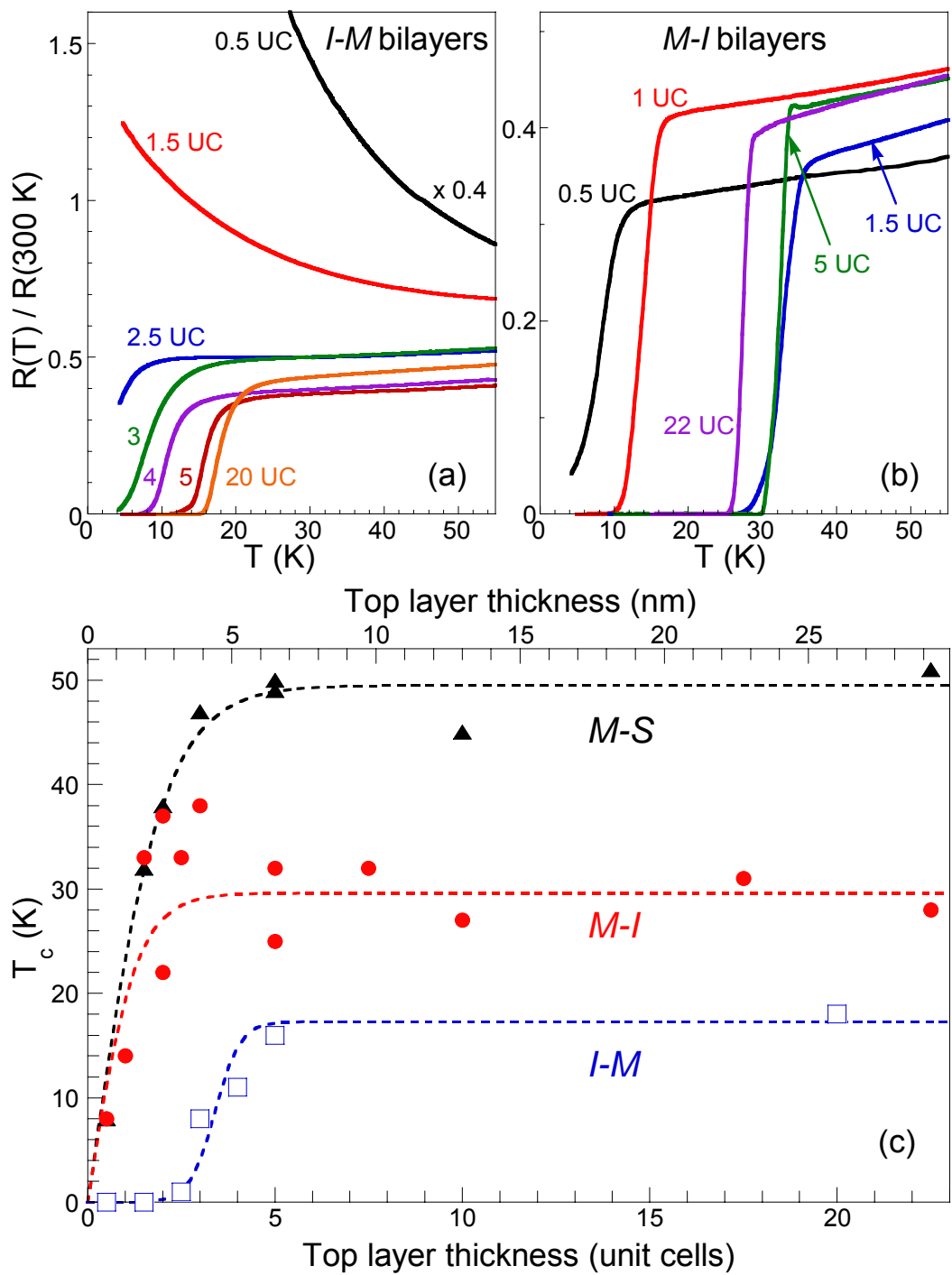

**Figure 2**

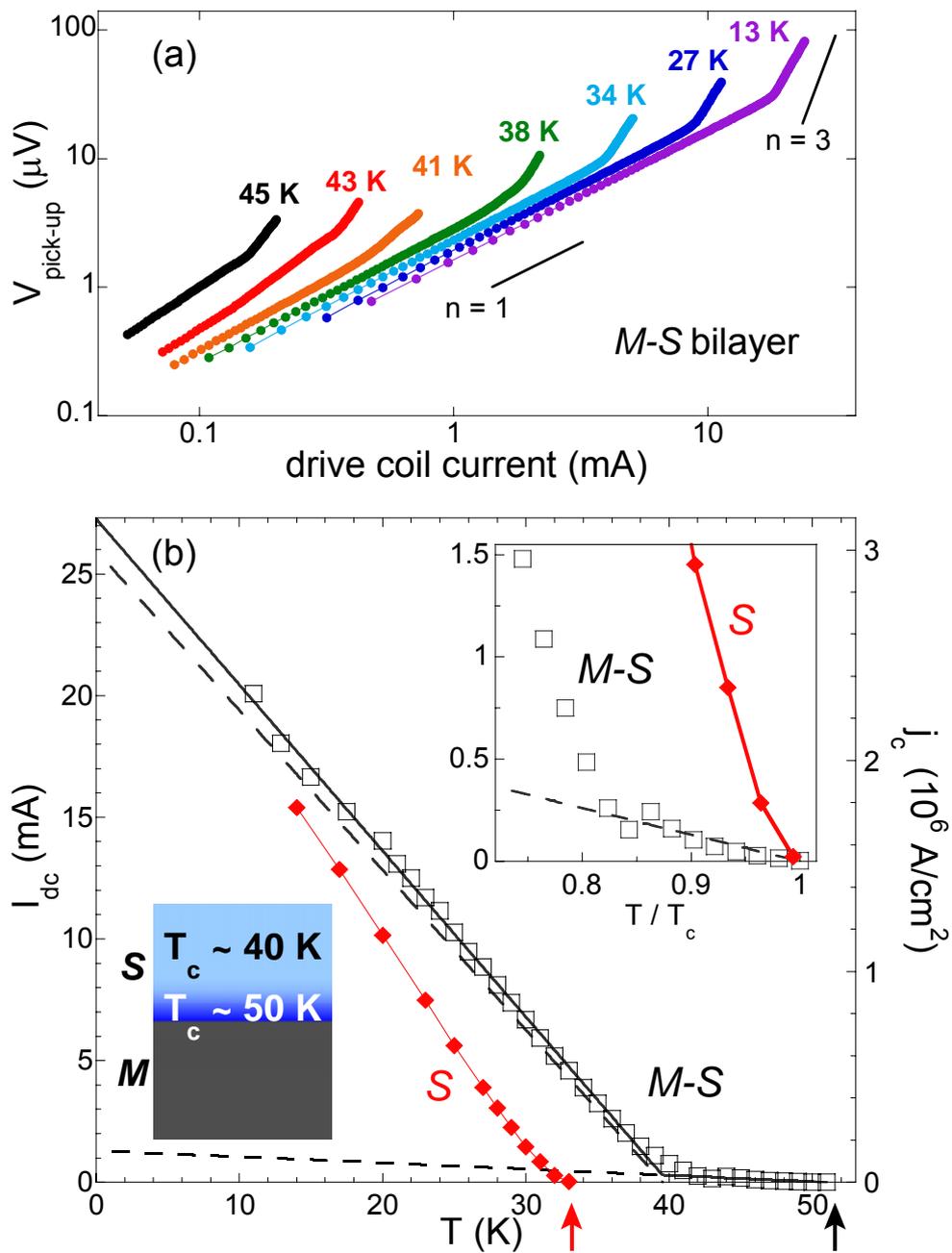

**Figure 3**

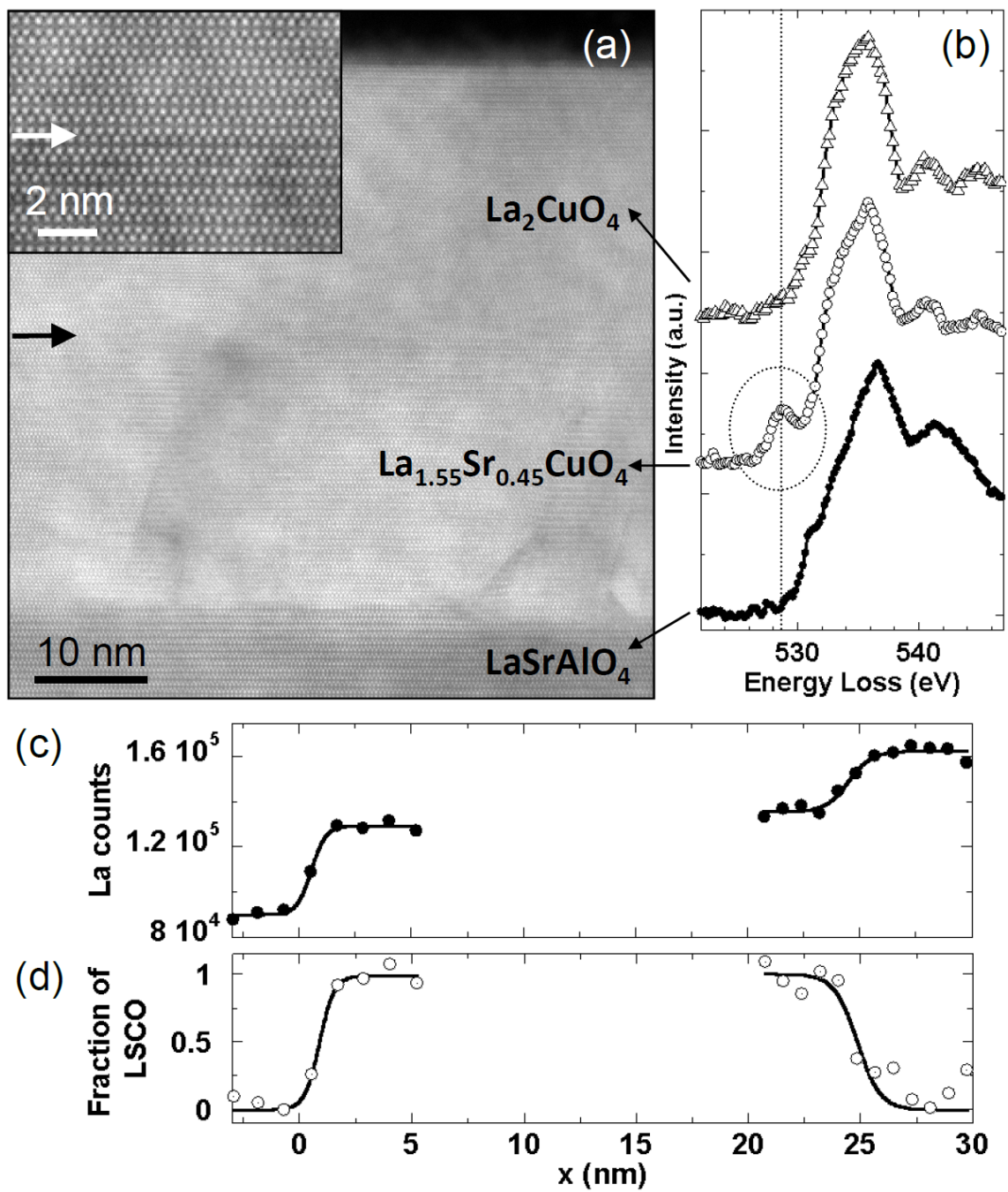

**Figure 4**

# Supplementary Information for "High-temperature interface superconductivity between metallic and insulating cuprates"

**A. Information about film growth and transport measurements**

The $La_{2-x}Sr_xCuO_4$ (LSCO) films were grown by atomic layer-by-layer molecular beam epitaxy on 10x10 mm$^2$ LaSrAlO$_4$ (LSAO) substrates. Resistivity measurements were made in the four-point-contact configuration with the current and the voltage leads wire-bonded onto evaporated Au pads. Mutual inductance data were acquired at a frequency $\nu$ = 10 kHz using a lock-in amplifier in the 'transmission' geometry - the sample was sandwiched between the drive and pick-up coils. The average radius of coils is 0.9 mm. At each temperature, the value of critical current was determined as the point above which the correlation function for the low current linear fit dropped below R = 0.9995.

**B. The highest critical temperature in single-phase films**

In Supplementary Fig. 1 we show a $T_c$ histogram of about 150 single-phase LSCO films grown in our laboratory. We varied widely the Sr doping level (spanning the entire phase diagram) as well as the film thickness and its oxygenation state. We have never seen $T_c$ exceeding 40 K in any of these films, in contrast to $T_c \sim$ 50 K obtained reproducibly in *M-S* bilayers. These data suggest that the 25% increase in $T_c$ might originate from the interface. The mutual inductance measurements (Fig. 3 of the manuscript and Supplementary Fig. 6) confirmed this observation. The same arguments make the Sr inter-diffusion mechanism unlikely to explain this relative enhancement: it seems improbable that whenever we grow a bilayer we always achieve (by uncontrolled inter-diffusion) the optimal Sr concentration for $T_c \approx$ 50 K, but we always (150 times) miss it whenever we grow a single layer.

**C. Other Information about La/Sr inter-diffusion at the interface**

We list below additional experimental observations in connection to possible La/Sr inter-diffusion across the interface. While arguably not as conclusive as the Scanning transmission electron microscopy (STEM) results shown in the main manuscript, these results support and are fully consistent with our conclusion that the cation inter-diffusion mechanism cannot be responsible for the interface effects reported here.

**C1. Reflection high-energy-electron-diffraction (RHEED) data.** Real time grazing-angle electron diffraction provides atomic-scale information about smoothness of the surface. It can reveal nucleation of secondary-phase precipitates that could emerge because of inaccurate stoichiometry or inadequate thermodynamic parameters during growth. Oscillations of the intensity of the specular spot in RHEED are a well-established signature of atomically smooth layer-by-layer growth. Furthermore, every compound displays its own pattern of amplitude and shape of RHEED oscillations because of characteristic form-factors determined by the chemical composition, the nature of the surface states (for instance metallic or insulating), or the specific growth mode at a given temperature and pressure. RHEED is thus an excellent tool to get qualitative information about transitions from one type of layer to another and whether this is done in a continuous or an abrupt fashion.

A typical pattern of RHEED oscillations recorded during growth of an ...-*I-M-I-M-*... superlattice is shown in Supplementary Fig. 2. Whenever we switch between the two materials, the pattern (both the amplitude and the shape) of the oscillations changes abruptly on the 0.5 UC scale from the one typical of single-phase *I* films to the one typical of single-phase *M*. This indicates that the interfaces are atomically sharp with respect to the cation composition, irrespective of the deposition sequence, and argues against a massive Sr inter-diffusion over more than 0.5 UC thickness in the growth direction.

**C2. Low energy ion scattering data.** Our MBE deposition chamber is equipped with a Time-of-flight ion scattering and recoil spectroscopy (TOF-ISARS) system, a surface-sensitive technique for *in-situ* measurements of the chemical composition[28]. It allows us to set an absolute upper limit on the amount of possible Sr diffusion along the growth direction. In Supplementary Fig. 3 we show the evolution of the peak associated with recoiled Sr from the top surface layers as a result of elastic binary collisions with the incoming 10 keV $K^+$ projectiles. The parameters were tuned to maximize the surface sensitivity: we used a low incidence angle ($\alpha$ = 15°), a low-index crystallographic azimuth, [100], and monitored single-scattering events.

Assuming that the integrated intensity of the Sr recoil peak is proportional to the surface concentration of Sr (open symbols in Supplementary Fig. 3), we can put an upper limit of 1 UC for the scale over which Sr diffusion could provide a doping level comparable to the one in LSCO with $T_c$ ~ 30 K (ref. 29). This is an overestimate because a substantial contribution in the TOF-ISARS spectra comes from projectiles that penetrate beyond the top 0.5 UC thick layer. Some residual Sr intensity can even be picked up from outside of the substrate area when the $K^+$ beam is swept across the sample. To obtain a rough estimate of the Sr profile we assume that scattering could arise from the 1.5 UC thick top layer and that possible cation inter-diffusion is proportional to the nominal difference in the Sr concentration between adjacently deposited layers. This model reproduces very well the experimentally determined intensities and, as expected, reveals indeed a more abrupt Sr profile (solid symbols in Supplementary Fig. 3) which can be identified as the LSCO fraction across the interface. Note that this distribution is in good agreement with the STEM results shown in Fig. 4 of the manuscript.

**C3. Transport in *I-M-I* trilayers.** In *M-I* bilayers, a typical $T_c$ is around 30 K, as seen in Fig. 2(c) of the manuscript. In contrast, *I-M-I* trilayers have $T_c$ reduced by at least 15 K, see Supplementary Fig. 4. The only difference between the *M-I* interfaces in the bi- and trilayer devices is the nature of the bottom layer and the proximity to the LSAO substrate. Neither cation inter-diffusion nor oxygen off-stoichiometry alone can explain this result. In Supplementary Fig. 4, we compare a 35-12 UC thick *M-I* bilayer with a 35-35-12 UC thick *I-M-I* trilayer. One could speculate that perhaps in the later film the top *M-I* interface got rougher because it was grown on a twice as thick buffer layer and thus some defects may have accumulated. However, we saw no signature of surface degradation using our real time in-situ growth monitoring tools. Moreover, we also grew *I-M-I* trilayers where the total thickness of the bottom *I* and *M* layers combined was twice smaller, i.e. 35 UC (e.g. 23-12-12), and also observed a $T_c$ reduced by 15-20 K. These findings suggest that Coulomb interactions and lattice effects are responsible for this trilayer effect[30,31].

**D. Analysis of the mutual inductance measurements**

In Supplementary Fig. 5, we show the temperature dependence of the mutual inductance in the single-phase and bilayer samples discussed in Fig. 3 of the manuscript. The sharp drop in the inductive response shows that $T_c$ = 33.2 K and $T_c$ = 51.6 K in the two samples, respectively. The data in Supplementary Fig. 5 were acquired with the drive coil current $I_d$ = 5 µA. Critical current measurements were performed at fixed temperature and varying the drive coil current. The slope of the linear response in the data, see Fig. 4a and Supplementary Fig. 6, renders the value of the mutual inductance at that temperature.

Following the method outlined in ref. 18 the peak value of the screening current at a fixed temperature was estimated starting from the integral equation:

$$-\mu_0 \lambda^2 \vec{j}(\vec{x}) = \frac{\mu_0}{4\pi} \int d\vec{x}' \frac{\vec{j}_d(\vec{x}')}{|\vec{x}-\vec{x}'|} + \frac{\mu_0}{4\pi} \int d\vec{x}' \frac{\vec{j}(\vec{x}')}{|\vec{x}-\vec{x}'|}$$

and solving it by the Fourier transform. Here $j_d$ and $j$ are the current densities in the drive coil and the film, respectively, and $\lambda$ is the London penetration depth. For a single drive coil the average current through the thin film is given by:

$$j(r) = -\frac{I_d R_d}{d\,\lambda^2} \int_0^\infty dq\, \frac{q}{Q^2(q,\lambda)}\, J_1(qR_d) \cdot J_1(qr)\, \frac{exp(-qD_1)}{1 + \frac{q}{Q}\frac{Cosh(Qd/2)}{Sinh(Qd/2)}}$$

Here $r$ is a radial distance in the film, $I_d$ and $R_d$ are the drive coil current and radius, respectively, $D_1$ is the distance from the film to the drive coil, $d$ is the film thickness, $Q^2 = q^2 + 1/\lambda^2$, and $J_1(x)$ is the 1$^{st}$-order Bessel function. The result for an array of drive coils preserving azimuthal symmetry like in our experimental setup is obtained by summing the individual contributions of single loops. This formula is valid in the linear regime when the screening is proportional to the current in drive coil. This is no longer true above certain values of $I_d$, when the current density $j$ in some regions of the film reaches the critical value[17,32].

The critical current density, $j_c$ (left scale of Fig. 3b), was identified with the maximum value of $j(r)$ corresponding to the experimentally determined 'critical' drive coil current, $I_{dc}$, see the right scale of Fig. 3b and Supplementary Fig. 6. A typical value of the zero temperature penetration depth, $\lambda_0 \sim 2{,}000$ Å, and an empirical temperature dependence, $\lambda(T) = \lambda_0 / [1 - (T/T_c)^4]^{1/2}$, were employed. Note that because of the exponential factor in the integrand the estimated value of $j_c$ is quite insensitive to the chosen parameters for $\lambda(T)$ in the limit $\lambda^2 \ll d \cdot D_1$. This condition is well satisfied in our films because $d \cdot D_1 \approx 1.5 \times 10^9$ Å$^2$ up to temperatures very close to $T_c$. The values obtained from the onset of non-linearity in the mutual inductance data agree well with those determined from the appearance of the 3$^{rd}$ harmonic in the pick-up voltage, as well as from direct measurements of *I-V* characteristics in a typical four-probe resistance configuration.

**E. Other information about the nature of interface superconductivity**

**E1. Eliminating effects of variability in growth and cool-down process.** To rule out possible effects of variability in growth and cool-down processes, we have devised two methods of producing single-phase layers and bilayers simultaneously. First, we have grown *I* (*M*) layers concurrently on bare substrates and on previously grown *M* (*I*) films. Second, we used ion-milling to remove the top layer from a part of the bilayer film, patterned in such a way that we could measure the two parts independently. In either case the result is that bilayers are superconducting while single-phase films are not. These data directly tie superconductivity to the interface between metallic and insulating cuprate layers. We have also performed similar control experiments to compare *M-S* structures with single-phase *S* films grown simultaneously and indeed found that the bilayers had $T_c$ systematically higher by at least 10 K. For the case of *I-M* and *M-I* bilayers the samples were vacuum annealed *in situ* on cool down from about 550° C. Further prolonged *ex situ* annealing did not change $T_c$ indicating that we are in the regime where there is no bulk interstitial oxygen while at the same time the $CuO_2$ plane structure remains intact.

**E2. Asymmetry between the *M-I* and *I-M* bilayers.** The origin of the asymmetry in the superconducting properties of *I-M* and *M-I* bilayers can be inferred from the resistivity data shown in Supplementary Fig. 7. As the thickness of the *M* layer is increased in *I-M* structures, the conductance at a fixed temperature stays low up to 2.5 - 3 UC and then it crosses over to a uniform linear increase. One possibility is that the first few *M* layers are disordered. In *M-I* structures, superconductivity occurs even when the top *I* layer is just 0.5 UC thick (Fig. 2a). The arguments so far suggest that the difference between *M-I* and *I-M* interfaces is in the presence of a 'dead' layer between the individual components in *I-M* bilayers. This conclusion is further corroborated by other experimental observations: the higher $T_c$ in *M-I* bilayers, the broader superconducting transition width in *I-M* structures and the absence of enhanced $T_c$ in *I-M* bilayers after ozone annealing.

A factor which could be the cause of the disordered barrier and the presence of dead layers is oxygen non-stoichiometry. Note first that the LSAO substrate and the LSCO cuprate films in this study are stacks of polar layers. For example in $La_2CuO_4$ $[CuO_2]^{2-}$ planes alternate with two successive $[LaO]^+$ layers. At both the substrate-cuprate and *M-I* interfaces there is a change in the layer charge alternation pattern. In order to minimize the mismatch in polarization and/or chemical potential, the top *I* layer near the *M-I* interface can trap interstitial oxygen during growth and become metallic and superconducting. For the same reasons one could expect that the *M* layers next to *I-M* interfaces to lose oxygen from $CuO_2$ planes, thus becoming disordered, localized and insulating.

Further analysis of the transport data in *I-M* case along these lines shows that for 1.5 UC thick top *M* layer the $R(T)$ dependence shows a minimum around 75 K and a ratio $R(300K) / R(4K)$ close to unity, nearly the same as what is observed in underdoped LSCO crystals with x $\approx$ 0.03-0.04 (ref. 33). The resistivity in such crystals is $\rho \approx$ 1 mΩ·cm; if it were the same in our underdoped layer, its thickness should be $d \approx$ 2 nm (~ 1.5 UC). These results also indicate that in *I-M* structures with a very thin top *M* layer the latter is actually underdoped and disordered. The normal structure of the *M* layer is recovered, however, after the critical thickness of 2-3 UC.

Finally, one should be aware that both the oxygen non-stoichiometry scenario described above as well as any possibly different local structures at the *I-M* and *M-I* interfaces will directly impact the electro-chemical potential and the dielectric properties. This can cause asymmetry in the screening length (and implicitly in the superconducting properties) at the two interfaces, see also section E4.

**E3. Oxygen non-stoichiometry and $T_c$ enhancement.** Several preliminary observations suggest that in *M-S* bilayers the enhanced $T_c \sim 50$ K is confined to the interface and also that this effect is tied to the way the interface affects incorporation of interstitial oxygen. The $T_c = 50$ K plateau is reached with 3 nm thick top *S* layer (see Fig. 2c) and it does not change with further thickness increase of the top layer. This shows that this is the characteristic length scale for the enhanced superconductivity. Clear evidence that in *M-S* bilayers interstitial oxygen acts differently than in single-phase films comes from annealing experiments: vacuum annealing of a 40 UC thick *S* film at $T \sim 200°$ C for 30 min. converts this compound from a superconducting metal into a strong insulator[5]. In contrast, in an *M-S* bilayer the critical temperature drops by only $\Delta T_c \approx 2$-3 K even if it is twice thinner, annealed four times longer (120 min.) and at higher temperatures (250°C). Interstitial oxygen in $La_2CuO_{4+}$ is mobile and, in particular in very thin films, it diffuses out of the sample on the scale of hours or days. Interface trapping of oxygen could be the reason for remarkable resilience and stability, over time scales of years, of the enhanced interface superconductivity in the *M-S* case.

**E4. The charge-transfer mechanism.** We consider also charge accumulation depletion due to a difference in the chemical potential. As mentioned in the manuscript, previous X-ray photoemission data[25] indicate that there is essentially no change in the chemical potential ( ) up to the optimal doping in LSCO, i.e. $d\mu/dx < 0.2$ eV/hole for $x \leq 0.16$. However, at higher doping, a larger decrease ($d\mu/dx \approx 1.5$ eV/hole) is observed. These results are consistent with the absence of a supercurrent when *I* layers are sandwiched between blocks of optimally doped LSCO, implying chemically abrupt interfaces and no inter-diffusion, while allowing for charge accumulation/depletion in *...-M-I-M-I-...* superlattices. Direct evidence for this effect has recently been reported in

ref.23. That study also addresses quantitatively the issue of inter-diffusion and the results agree very well with the present STEM data.

Assuming a carrier density $n \approx 4.8 \times 10^{21}$ cm$^{-3}$ for $M$ layers, $\varepsilon_r \sim 30$ (ref. 34) for the dielectric permitivity of $I$ and $\Delta\mu \sim 0.5$ eV (ref. 24), the formula for the accumulation layer at a metal-semiconductor interface, $\Delta\mu = e\, n\, l^2 / 2\, \varepsilon_r\, \varepsilon_0$ (ref. 35) gives a value $l \approx$ 0.6 nm, or about 0.5 UC. Interestingly, this crude estimate falls within the range of the characteristic length for interface superconductivity determined from Fig. 2 and is in quantitative agreement with the screening length at the $M$-$I$ interface determined in ref. 23.

**F. Electron microscopy: experimental details and information about interface roughness**

The specimen was prepared using the focused ion beam in-situ lift-out technique on an FEI Strata 400S DualBeam instrument[36]. The specimen was polished using 2 keV Ga$^+$ ions as the final step to facilitate the STEM imaging[37]. The electron microscopy and spectroscopy measurements were performed on a monochromated 200 kV FEI Tecnai F20-ST STEM with a minimum probe size of ~1.6Å and a convergence semiangle of (9.5±1) mrad. The ADF image was recorded with a detector inner angle of ~65 mrad. To increase signal to noise and average out the scan noise, 10 successive images (5 for the lower magnification ADF image shown in Fig. 4 in the main text), each recorded at 8 microseconds per pixel, were cross-correlated and averaged. Subsequently, the 1024×1024 pixel images were rebinned to 512×512 pixels.

For the EELS measurements, the Tecnai F20-ST is equipped with a Gatan imaging filter 865-ER. The energy resolution was ~0.6 eV as measured from the FWHM of the zero loss peak at 0.2 eV/channel. The La-M$_{4,5}$ and O-K spectra were recorded simultaneously on a 2048×2048 pixel CCD. The O-K edges, shown in Fig. 4(b), were recorded for 60 s in the respective layers. To increase signal to noise in the line profiles across the substrate-M and the M-I interfaces, each spectrum in the series was recorded for 20 s with the electron beam scanning over a 20-30nm long line parallel to the interface. While this averaging improves the signal to noise and prevents radiation damage, it may contribute to the obtained interface root mean square (rms) width. An additional contribution to the rms width at the *M-I* interface may come from the projection of extended anti-phase/out-of-phase defects observed in the *M* layer which nucleate at the cuprate-substrate interface (Supplementary Fig. 8) and are most likely due to local variations in the termination layer of the substrate. Hence, the obtained rms interface roughness, $\sigma$ = 1.2±0.4 nm ( $\leq$ 1 UC), at the *M-I* interface sets an upper limit to any cation intermixing.

**Supplementary references**

**Supplementary figures:**

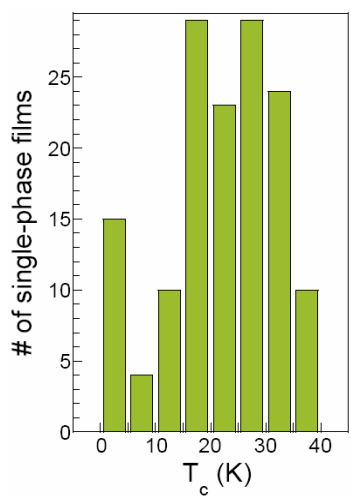

**Supplementary Figure 1**: Histogram showing the values of $T_c$ of the single-phase LSCO films grown by molecular beam epitaxy at Brookhaven National Laboratory. The film thickness spans the whole range discussed in our work and the Sr concentration varies across the entire phase diagram. The films were also annealed under varying

conditions and characterized at different degrees of oxygenation, but the $T_c$ = 40 K limit was not exceeded.

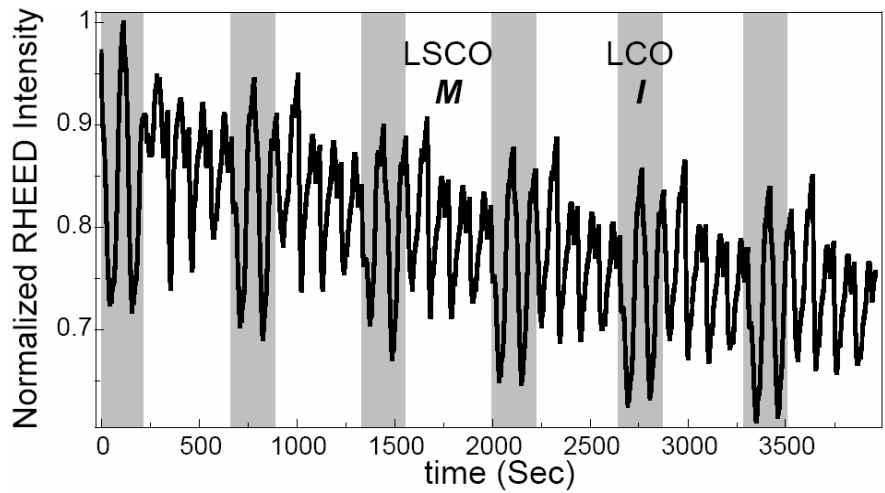

**Supplementary Figure 2**: RHEED intensity oscillations of the specular beam during growth of a [(1x*I*):(2x*M*)]$_n$ superlattice. The oscillation pattern changes discontinuously in both the shape and the amplitude between *I* and *M* layers, indicating that crossover from the insulating to the metallic layer is abrupt on the 0.5 UC scale.

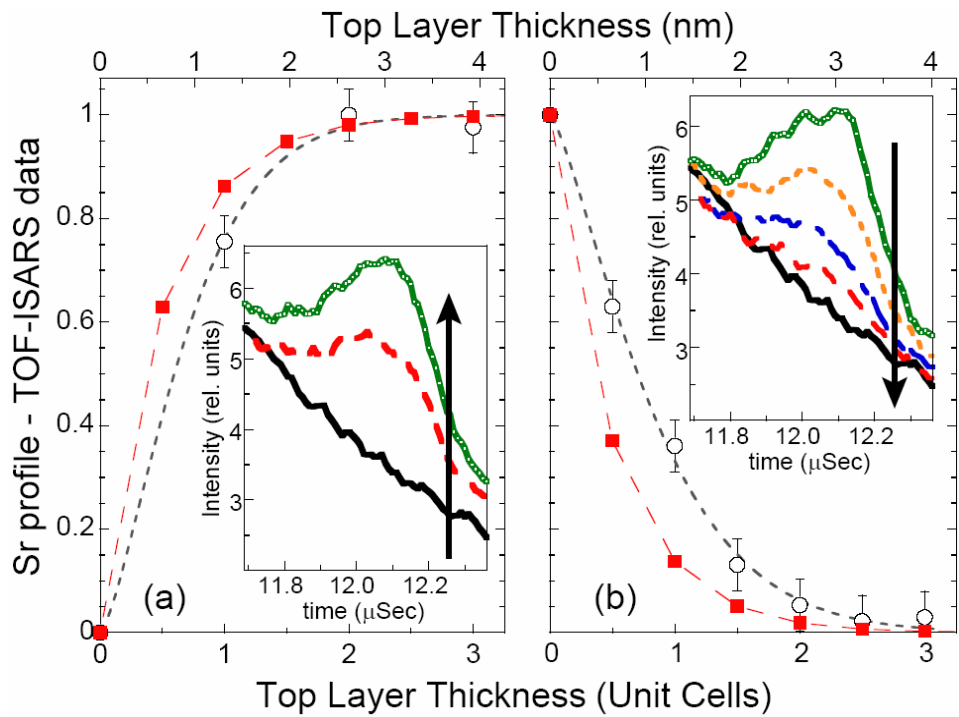

**Supplementary Figure 3**: (a) Time-of-Flight Ion Scattering and Recoil Spectroscopy (TOF-ISARS) data as a function of film thickness on *M* side of *I-M* structure. The main panel displays the normalized integrated intensity of the Sr recoil peak in *M* (open circles). The Sr concentration profile, estimated as described in the text, is displayed by solid squares. Dashed lines are guides for the eye. The inset shows the evolution of the Sr recoil peak. (b) Same as (a), but on *I* side of *M-I* structure.

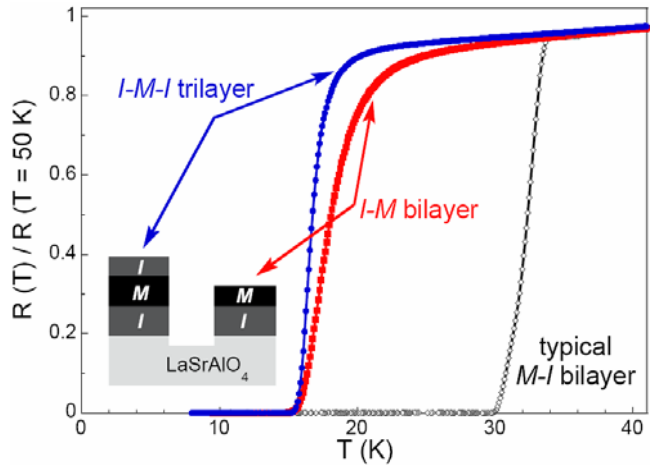

**Supplementary Figure 4**: The temperature dependence of normalized resistance in a 35 UC – 35 UC – 12 UC thick *I-M-I* trilayer film (solid blue circles) and in a bilayer (red solid squares) patterned from the same structure as shown in the inset. In *I-M-I* trilayers we observe the same $T_c$ (~ 15 K) as in *I-M* bilayers. Because of the presence of the bottom *I* layer, the top *M-I* interface in *I-M-I* structure has a $T_c$ reduced compared to the typical values (~ 30 K) in bilayers where *M* is adjacent to the substrate; a typical 40 UC – 5 UC *M-I* bilayer data are shown by open black diamonds.

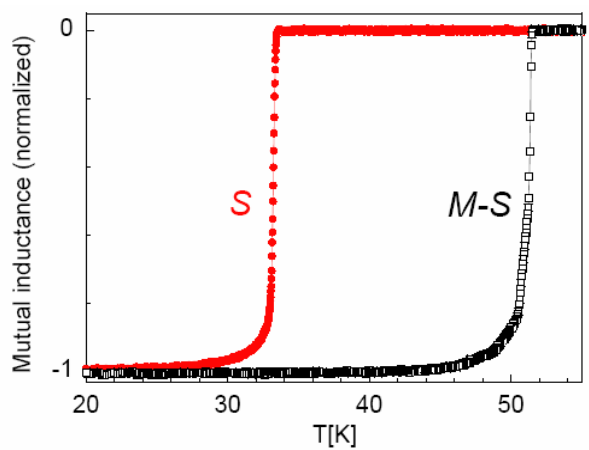

**Supplementary Figure 5**: The temperature dependence of the imaginary part of the mutual inductance of the single phase, *S*, and bilayer, *M-S*, samples discussed in Fig. 4 of the manuscript. The onset of the superconducting Meissner response occurs at $T_c$

of each sample. The sharp drop in the imaginary component indicates a very narrow superconducting transition.

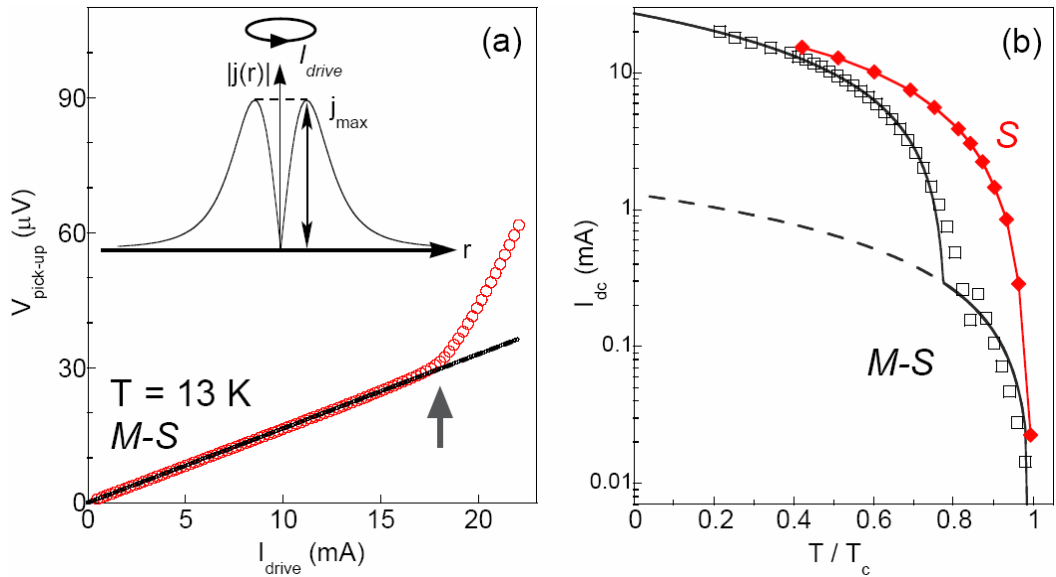

**Supplementary Figure 6**: (a) Typical dependence of the pick-up voltage on the drive coil current, $I_{drive}$, at a given temperature in our superconducting films. The data shown here are taken from a *M-S* bilayer at T = 13 K. The arrow marks the value of the critical drive coil current, $I_{dc}$, i.e. the deviation from linearity. The inset shows a typical radial current density profile *j(r)* screening the magnetic field generated by $I_{drive}$. The current profile was calculated as described in section (D) of the text. The critical current density in the film was identified with $j_{max}$, the peak value of the screening current corresponding to $I_{dc}$. (b) The dependence of $I_{dc}$ in *M*-S and *S* films on $T/T_c$ in the whole temperature range. The data are the same as in Fig. 3b of the manuscript but here the interface contribution in the *M-S* case (dashed line) is emphasized by the semi-logarithmic scale.

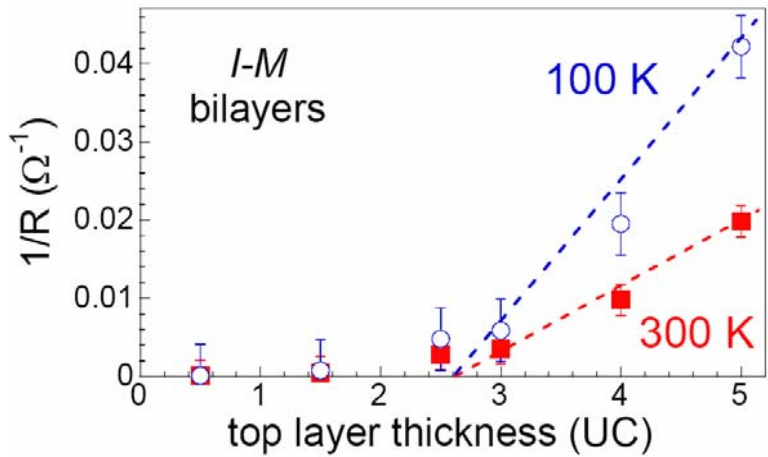

**Supplementary Figure 7**: The dependence of the inverse resistance on the thickness of top layer in *I-M* bilayers at T = 100 K (empty circles) and T = 300 K (filled squares), respectively. The thickness of the bottom insulating layer in all cases was 40 UC. The dashed lines are guides for the eye.

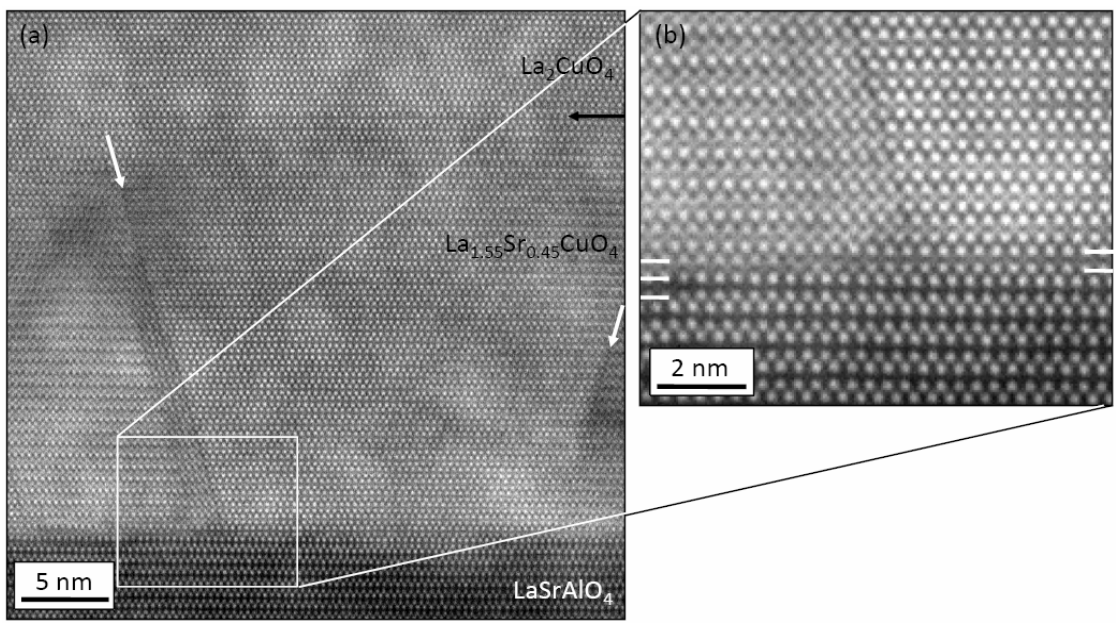

**Supplementary Figure 8**: (a) Annular dark field image of the structure showing extended defects in the *M* layer (marked by white arrows). The black arrow shows the *M-I* interface (b) A magnified image of one defect which nucleated at the cuprate substrate interface and is due to local variations in the termination layer of the substrate.